\documentclass[doublecol]{epl2}

\usepackage{amsmath,amssymb}    
\usepackage{graphicx}   
\usepackage{verbatim}   
\usepackage{color}      

\title{Excitability in autonomous Boolean networks}

\author{David P. Rosin\inst{1,2}\thanks{E-mail: \email{rosin@phy.duke.edu}} \and Damien Rontani \inst{1} \and Daniel J. Gauthier\inst{1} \and Eckehard Sch\"oll\inst{2}}
\shortauthor{D. Rosin \etal}
\institute{                    
  \inst{1} Department of Physics, Duke University, Durham, North Carolina 27708, USA\\
  \inst{2} Institut f\"ur Theoretische Physik, Technische Universit\"at Berlin - Hardenbergstr. 36, D-10623 Berlin, Germany, EU
}

\pacs{05.45.-a}{Nonlinear dynamical systems}
\pacs{87.19.lr}{Control theory and feedback in neuroscience}

\abstract{ 
We demonstrate theoretically and experimentally that excitable systems can be built with autonomous Boolean networks. Their experimental implementation is realized with asynchronous logic gates on a reconfigurabe chip.
When these excitable systems are assembled into time-delay networks, their dynamics display nanosecond time-scale spike synchronization patterns that are controllable in period and phase.
}
\begin{document}

\maketitle

\section{Introduction}
Excitability is a property of dynamical systems, where the system rests in a stable
fixed point, but large excursions in phase space ({\em spikes}) can be generated in response to small perturbations above a threshold \cite{IZH07}. When coupled in networks, excitable systems can exhibit complex spatio-temporal spike patterns and synchronized oscillations as observed in chemical reactions \cite{STE93a}, heart tissue \cite{DAV92a}, and populations of interacting neurons \cite{VAR12,KEE09}. This abundance of excitability in nature has motivated many theoretical and experimental studies.

Theoretical approaches, such as the paradigmatic model for excitability proposed by Fitz\-Hugh and Nagumo \cite{FIT55,NAG62}, have helped to uncover and understand the diverse collective behaviors (bursting, cluster synchronization, and phase transitions, for example) that arise in networks of excitable systems \cite{DAH12,SCH08,LEH11,KAN11a,VIC08}. However, they usually do not fully integrate all experimental imperfections and heterogeneities like noise and system parameter variation, which may have significant impact on the dynamics.

This motivates experimental studies on excitable systems, for example, with analog electronic circuits. These electronic systems can be realized in very large scale integration (VLSI), where up to thousand excitable systems are implemented on a custom analog chip and the coupling topology is handled by a separate reconfigurable digital chip \cite{IND11,ART06}. This configuration, however, presents major hindrances due to speed limitations, the cost and long design cycle time of the custom analog chip, and its connections to the digital reconfigurable chip. Of particular concern is the fact that the analog signal is digitized, leading to discretization errors in the coupling.

To address these issues, we propose an excitable system built from continuous-time asynchronous logic gates. This system is based on autonomous time-delay Boolean networks that have been found previously to show oscillatory dynamics and chaos depending on the choice of topology and Boolean functions \cite{ZHA09a,CAV10,MES97,GLA98,EDW00,GHI85}. With our approach, we can experimentally realize the excitable systems  and couple them to networks on a single inexpensive field-programmable gate array (FPGA).

The resulting dynamics are spike synchronization patterns reproducible with a theoretical Boolean map. They are controllable by the network's link delay times \cite{PAN12} and are much faster than the dynamics of common networks of excitable systems. 
The nanosecond timescales give our approach an advantage for potential ultra-fast neuro-inspired data processing \cite{sch08k}, similar to reservoir computing \cite{JAE04}.

\section{Experimental setup}
Our design realizes three important properties characterizing excitability~\cite{IZH07}: (i)~the all-or-none principle, where the system responds only if an input is above a threshold and stays quiescent otherwise; (ii)~pulse dynamics, where output pulses have fixed width independent of the input pulse shape; and (iii)~a refractory phase, where the excitable system remains unresponsive during a refractory period after generating an output pulse.

Asynchronous logic gates are well suited to fulfill these properties. The all-or-none principle~(i) is intrinsically embedded in logic gates \cite{MCC43}. Their output voltages $V$ transition between $V=V_\mathrm{high}$ (the \textit{high} level) and  $V=V_\mathrm{low}$ (the \textit{low} level) as their inputs cross a threshold $V_\mathrm{th}$. Pulse dynamics~(ii) and the refractory phase~(iii) can be realized through pulse generators (PGs), which exploit the intrinsic propagation delays of logic gates $\tau_\mathrm{gate}$.

\begin{figure}[tb]
\centering
\includegraphics[width=\linewidth]{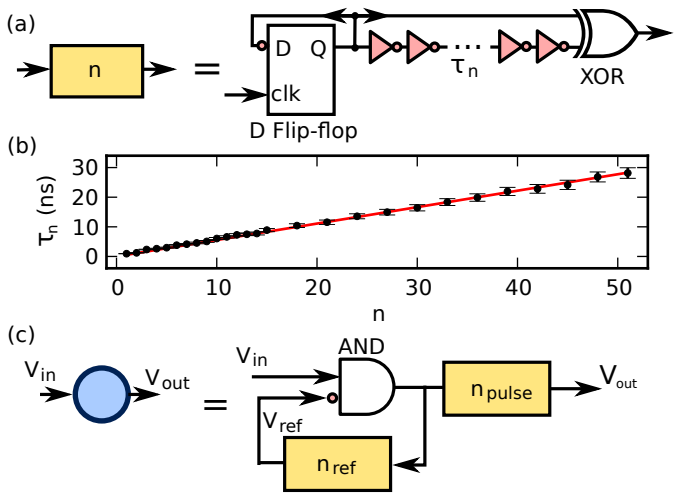}%
\caption{(Color online) (a) Setup of the pulse generator (PG) characterized by an integer $n$, which represents the number of pairs of inverters in the delay line and thus its time delay $\tau_n\sim 2n\tau_\mathrm{gate}$. The pairs of inverters act as time delays that do not change the Boolean state. (b) Experimental measurement of the delay time $\tau_n$ comprising $n$ pairs of inverter gates, with relative error of $\sim\pm 3.5\%$ (error bars) and linear regression (red line) with $\tau_n\sim2n\cdot 280\un{ps}$. (c) Complete design of an excitable node; it combines one pulse generator labeled $n_\mathrm{pulse}$ to realize the pulse dynamics with another one labeled $n_\mathrm{ref}$ to realize the refractory phase. The refractory period of the excitable system $T_\mathrm{ref}$ and its pulse width $T_\mathrm{pulse}$ are determined by the integers $n_\mathrm{ref}$ and $n_\mathrm{pulse}$, respectively. The voltages $V_\mathrm{in}$ and $V_\mathrm{ref}$ are inputs to an AND gate, where the second input is inverted, as indicated by a circle.}
\label{fig:setup_artificial_neuron}
\end{figure}

As shown in Fig.~\ref{fig:setup_artificial_neuron}a, the PG is implemented using a D-type flip-flop, a delay line of duration $\sim\tau_n$ made of $2n$ inverters with $n\in \mathbb{N}$, and an asynchronous logic gate executing the XOR operation \cite{JEO04}. Its dynamics consists of the generation of a single pulse of width $T_n$ in response to a positive edge (low to high transition). More specifically, in response to a positive edge at its clock (clk) input, the flip-flop with connection from output (Q) to inverted input (D) generates a Boolean transition at its output (Q).
This signal reaches the XOR gate inputs with a time-delay difference $\sim\tau_n$, due to a delay line. As a consequence, the XOR gate has different input logic values during the time delay and hence generates a high voltage $V_\mathrm{high}$ of width $T_n=\tau_n\propto n$, with delay $\tau_n$ measured in Fig.~\ref{fig:setup_artificial_neuron}b.

We combine two PGs with an AND gate, as depicted in Fig.~\ref{fig:setup_artificial_neuron}c, so that the system exhibits excitable dynamics in its output voltage $V_\mathrm{out}$ in response to an above-threshold input voltage $V_\mathrm{in}$. The PG labeled $n_\mathrm{pulse}$ ($n_\mathrm{ref}$) produces a voltage pulse $V_\mathrm{pulse}$ ($V_\mathrm{ref}$) of width $T_\mathrm{pulse}$ ($T_\mathrm{ref}$). The voltage $V_\mathrm{ref}$ indicates whether or not the system is in its refractory phase ($V_\mathrm{ref}$ high or low, respectively); its interplay with $V_\mathrm{in}$ governs the dynamics of our excitable node.

When $V_\mathrm{ref}$ is low and $V_\mathrm{in}$ has a positive edge, the AND gate also generates a positive edge so that each PG produces a pulse. $V_\mathrm{pulse}$ is sent to the output of the excitable system and the high value of $V_\mathrm{ref}$ now blocks inputs $V_\mathrm{in}$ to the AND gate for the refractory period $T_\mathrm{ref}$ and, therefore, prevents pulse generation during that time. When the refractory phase ends, \emph{i.e.}, $V_\mathrm{ref}$ is back to a low voltage, the system becomes responsive to input excitations $V_\mathrm{in}$ again.

Our design of the excitable node is motivated by the dynamics of integrate-and-fire neurons \cite{BUR06}, where the membrane potential evolves as a function of its synaptic input. When inputs are present, the membrane potential increases (integration) until it reaches a threshold, the condition for generating a pulse (firing). In our approach, in contrast, the excitable system compares its input voltage directly to a threshold without an electronic analog of a membrane potential. 
Consequently, when increasing $V_\mathrm{in}$ above the threshold of our system, oscillations start with a constant (finite) period, so that our system exhibits a behavior analogous to type-II excitability \cite{IZH07}.
After generating a pulse, the membrane potential of integrate-and-fire neurons returns to a resting value and its dynamics is deactivated for a finite duration \cite{BUR06}, which is the same mechanism used in our system to realize a refractory period.

The implementation of our system can be realized with various technologies. Here, we use an FPGA because of the very large number of logic elements and flip-flops available (up to $\sim10^6$ \cite{MAX04}) and the possibility to operate them asynchronously. Another advantage of FGPAs is the re-programmability of the logic elements. They consist of CMOS-based multiplexers fed by rewritable memory bits with a specific arrangement to realize the logic gate operation. These multiplexers connect $N$ inputs ($N=4$ for the FPGA used here) to one output so that the overall behavior is that of a logic gate with the same operation. Finally, FPGAs ensure flexible connection of logic elements \cite{MAX04}, allowing us to realize large networks of excitable elements.

\section{Dynamics of one excitable node}

In this section, we conduct experiments on a single excitable node driven first by a constant input and second by self-feedback, constituting a simple network.
The experiments are realized on an Altera Cyclone IV FPGA (EP4CE115F29C7N), which has $\sim 115{,}000$ logic elements with  propagation time delay and rise time of $\tau_\mathrm{gate}=(280\pm10)\un{ps}$ and $\tau_\mathrm{rise}=(310\pm10)\un{ps}$, respectively. Signals generated within the FPGA pass through an additional input-output logic gate (hardwired to the output pins of the FPGA) before being acquired by a high-speed oscilloscope (DSO80804A) with 8 GHz bandwidth and 40 GSa/s sampling rate.
 
\begin{figure}[tb]
\centering
\includegraphics[width=\linewidth]{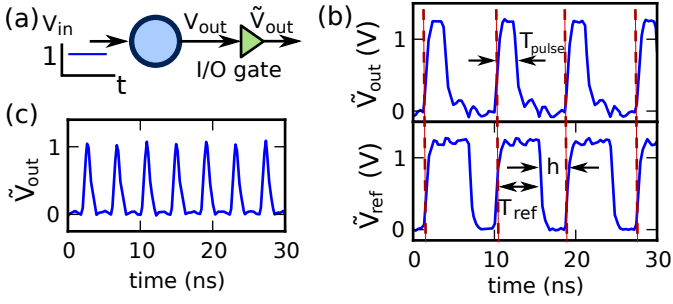}
\caption{(Color online) (a) An excitable node on the FPGA is subject to a constant above-threshold input voltage $V_\mathrm{in}$. The output voltage $V_\mathrm{out}$ and $V_\mathrm{ref}$ pass input-output (I/0) gates with outputs labeled $\widetilde{V}_\mathrm{out}$ and $\widetilde{V}_\mathrm{ref}$, respectively.
(b) Both voltages are recorded for $T_\mathrm{ref}=(5.40\pm0.05)\un{ns}$ ($n_\mathrm{ref}=10$), $T_\mathrm{pulse}=(2.34\pm0.05)\un{ns}$ ($n_\mathrm{pulse}=4$).
(c) Output $\widetilde{V}_\mathrm{out}$ of a minimal implementation of the excitable node with $T_\mathrm{ref}=(0.68\pm0.04)\un{ns}$ ($n_\mathrm{ref}=1$), $T_\mathrm{pulse}=(0.80\pm0.04)\un{ns}$ ($n_\mathrm{pulse}=1$).
}
\label{fig:1node_exp}
\end{figure}

We apply a constant, above-threshold input voltage $V_\mathrm{in}$ to the excitable node (Fig.~\ref{fig:1node_exp}a). In response, it generates periodic pulses with width of a few nanoseconds, as shown in Fig.~\ref{fig:1node_exp}b. In this regime, the system is oscillatory, similar to biological neurons with constant stimulus \cite{IZH07}. 
To understand the dynamics, we analyze the output voltage $V_\mathrm{out}$ and the voltage $V_\mathrm{ref}$ that indicates the refractory phase.

The pulses in $V_\mathrm{out}$ and $V_\mathrm{ref}$ are generated almost simultaneously at times indicated by vertical dashed lines in Fig.~\ref{fig:1node_exp}b. The pulse in $V_\mathrm{ref}$ indicates the refractory phase and its pulse width equals the refractory period. Therefore, the refractory phase starts at the dashed lines and ends when $V_\mathrm{ref}$ is low. Then, the system generates a new pulse, which is induced by a negative edge transition in $V_\mathrm{ref}$, since $V_\mathrm{in}>V_\mathrm{th}$. 
It requires an additional processing time $h$ to generate the output pulse, which is due to the flip-flops in the PGs and is measured to be $h=(3.2\pm0.4)\un{ns}$.
This processing time, together with the refractory period $T_\mathrm{ref}$, constitutes the period of the pulses in this experiment ($T=T_\mathrm{ref}+h$). The pulse width in $V_\mathrm{out}$ is given by $T_\mathrm{pulse}$.

The dynamics is determined by $T_\mathrm{ref}$, which accounts for the period of oscillations, and $T_\mathrm{pulse}$, which determines the width of output pulses of the system; these two quantities are controlled by parameters $n_\mathrm{ref}$ and $n_\mathrm{pulse}$, respectively, as shown in Fig.~\ref{fig:setup_artificial_neuron}b:
$T_\mathrm{ref}\approx 2n_\mathrm{ref} \tau_\mathrm{gate}$ and 
$T_\mathrm{pulse}\approx 2n_\mathrm{pulse} \tau_\mathrm{gate}$, as follows from the construction shown in Fig.~\ref{fig:setup_artificial_neuron}a. 
This can be seen experimentally when conducting the same experiment with different parameters $n_\mathrm{pulse}$ and $n_\mathrm{ref}$. For example, with  $n_\mathrm{pulse}=n_\mathrm{ref}=1$, which constitutes a minimal number of seven logic gates, the pulse widths $T_\mathrm{pulse}$ are on a sub-nanosecond scale and the period $T$ is dominated by $h$ (Fig.~\ref{fig:1node_exp}c).

Experimental fluctuations in $T_\mathrm{ref}$ and $T_\mathrm{pulse}$ are characterized in two ways. 
First, when comparing different measurements of $T_\mathrm{ref}$ and $T_\mathrm{pulse}$ on a single implementation, we obtain temporal fluctuations of $\pm 1\%$, with an origin described in the next paragraph. 
Second, when comparing the measurements of $T_\mathrm{ref}$ and $T_\mathrm{pulse}$ on several different copies of the same excitable node on the same chip, we obtain a significantly larger error of $\pm 3.5\%$. The origin of this error is heterogeneity in the propagation delays from logic element to logic element.

Finally, the dynamics of the excitable system is analog-like and fluctuates in pulse shape and timing. These non-ideal experimental behaviors originate from low pass filtering, jitter, and history- and state-dependency of the propagation time delays within logic gates \cite{CAV10,BEL00b}. The asynchronous logic operation is limited in speed only by the high-frequency cutoff of the logic gates and is much faster than common synchronous operation limited to $\sim100\un{MHz}$.

\begin{figure}[tb]
\centering
\includegraphics[width=\linewidth]{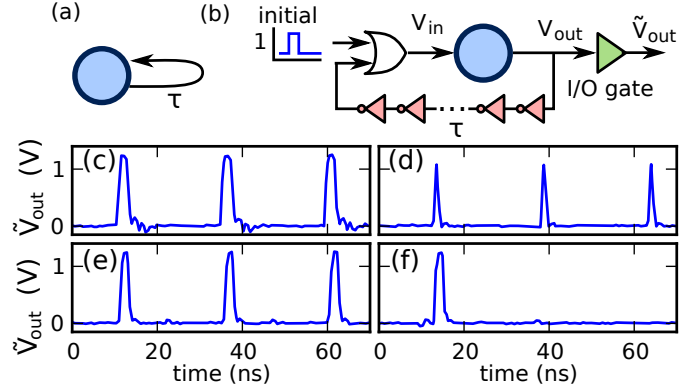}
\caption{(Color online) (a) Scheme of one excitable node with delayed feedback where the delayed feedback link is represented as an arrow. (b) Representation of (a) with logic elements used for the feedback. The pink triangles with circles represent 80 inverter gates that are incorporated to implement a time delay of $\tau=(21.3\pm0.5)\un{ns}$ ($n_\tau=40$). (c),(d) Resulting dynamics with parameters as in Fig.~\ref{fig:1node_exp}(b) and (c), respectively. (e) Same with $T_\mathrm{ref}=(10.75\pm0.05)\un{ns}$ ($n_\mathrm{ref}=20$), $T_\mathrm{pulse}=(1.95\pm0.03)\un{ns}$ ($n_\mathrm{pulse}=4$). (f) Same with $T_\mathrm{ref}=(24.04\pm0.05)\un{ns}$ ($n_\mathrm{ref}=45$), $T_\mathrm{pulse}=(2.05\pm0.05)\un{ns}$ ($n_\mathrm{pulse}=4$).}
\label{fig:1node_feedback}
\end{figure}

We also investigate the behavior of a simple network consisting of an excitable system with self-feedback with time delay $\tau$ (Fig.~\ref{fig:1node_feedback}a), \emph{i.e.}, $V_\mathrm{in}(t)=V_\mathrm{out}(t-\tau)$ \cite{FOS96}. The time delay accounts for non-instantaneous transmission times along network links (for example, the propagation time along nerve cells connecting different areas of the brain \cite{KEE09}).
 
The delayed feedback link is realized as shown in Fig.~\ref{fig:1node_feedback}b with $n_\tau=40$ cascaded pairs of inverter gates, each imposing its propagation delay to the path, leading to a total time delay of $\tau=(21.3\pm0.5)\un{ns}$, as characterized in Fig.~\ref{fig:setup_artificial_neuron}b. 
The delayed feedback signal and an initial stimulus are both applied to the excitable node. As an excitable node has only one input, an OR gate is used to combine the two signals. The OR operation allows both signals to excite the node, but also other logic gate operations to combine inputs are possible, as discussed later. Here, the system is operated in its excitable regime. 

When no initial stimulus is applied, the feedback system rests in a stable quiescent state.
But, when a pulse (with width $(1.6\pm0.1)\un{ns}$) is injected once, the system generates a periodic pulse train as shown in Fig.~\ref{fig:1node_feedback}c. Initializations with multiple pulses result in similar pulse trains with shorter periods.

The dynamics arise from the delayed feedback. When a pulse is generated by the system, it travels through the delay line during $\tau$. Then, it is input to the node to generate another output pulse after the processing time (system response time) $h$. Therefore, the period of the pulses is $T=\tau+h$ for this coupling scheme, which is confirmed by the experiment.

This behavior is reproduced for systems with parameters $n_\mathrm{ref}=n_\mathrm{pulse}=1$ that have a shorter refractory period and small pulse widths (Fig.~\ref{fig:1node_feedback}d) and for systems with parameters $n_\mathrm{ref}=20$ and $n_\mathrm{pulse}=4$ that have a longer refractory period (Fig.~\ref{fig:1node_feedback}e). However, when the refractory period is increased further to $n_\mathrm{ref}=45>n_\tau=40$, i.e., $T_\mathrm{ref} > \tau$,
self-sustained pulsing dynamics are no longer solutions of the system (Fig.~\ref{fig:1node_feedback}f). Instead, the system responds only to the initial stimulus and then stays in the quiescent state. 
When this first response is fed back after passing the delay line, the excitable node is still in the refractory phase and, therefore, cannot generate another output pulse.

\section{Two delay-coupled excitable nodes}
\begin{figure}[tb]
\centering
\includegraphics[width=\linewidth]{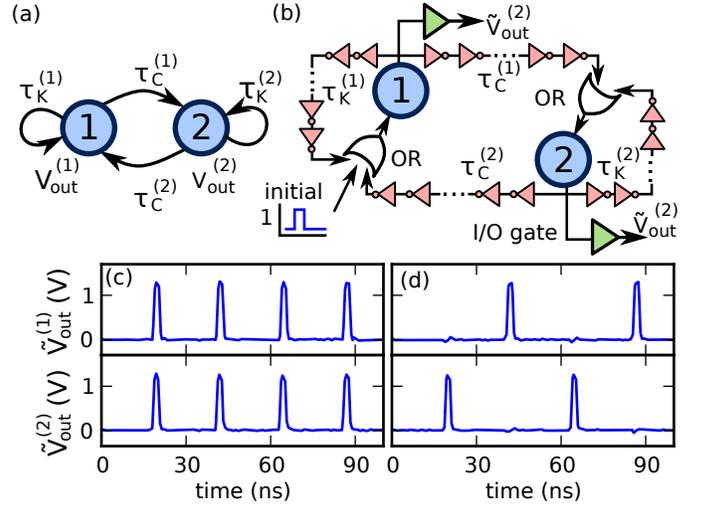}
\caption{(Color online) (a) Scheme of two excitable nodes with delayed mutual coupling and delayed feedback where delay links are represented as arrows. b) Representation with logic elements used for the coupling. The pink triangles with circles represent inverter gates that are incorporated in numbers $n_{\tau,C}$ and  $n_{\tau,K}$ to adjust the delay times indicated on the links. 
Node parameters are $n_\mathrm{pulse}=4$ ($T_\mathrm{pulse}=(2.1\pm0.2)\un{ns}$) and $n_\mathrm{ref}=10$ ($T_\mathrm{ref}=(5.3\pm0.2)\un{ns}$).
(c) Stable output of both nodes with coupling delays realized with $n_{\tau,C}=n_{\tau,K}=40$ pairs of inverters leading to link delays 
$\tau_C^{(1)}=(21.6\pm0.2)\un{ns}$, $\tau_C^{(2)}=(21.7\pm0.2)\un{ns}$, $\tau_K^{(1)}=(21.6\pm0.2)\un{ns}$, $\tau_K^{(2)}=(21.4\pm0.2)\un{ns}$. (d) Same as (c) with $n_{\tau,C}=40$, 
$\tau_C^{(1)}=(22.1\pm0.2)\un{ns}$, $\tau_C^{(2)}=(21.2\pm0.2)\un{ns}$, $n_{\tau,K}=80$, $\tau_K^{(1)}=(43.0\pm0.4)\un{ns}$, and $\tau_K^{(2)}=(43.5\pm0.4)\un{ns}$. Electrical cross talk between the node outputs is visible as small oscillations near the noise floor. Additional logic gates are used to measure the link delays of this specific implementation.}
\label{fig:2node_exp}
\end{figure}

In this section, we study a network of two delay-coupled excitable nodes with delayed feedback, as shown schematically in Fig.~\ref{fig:2node_exp}a. The coupling and feedback delays are denoted by $\tau_C^{(1,2)}$ and $\tau_K^{(1,2)}$, respectively. 

Here, we are interested in the parameter regime of synchronization between the two nodes, as such states are found to be important in the brain. For example, they help to understand cognition and learning and also pathological conditions such as Parkinson's disease \cite{SCH11e}.

The coupling scheme in Fig.~\ref{fig:2node_exp}a has already been studied with the Fitz\-Hugh-Nagumo model, and the parameter regimes where coherent oscillations appear and their period and phase have been found \cite{SCH08,PAN12,ERN95,SHA00}. These regimes form straight lines in the parameter space of the delay times, given by the following conditions. Without loss of generality, assume for the coupling delays $\tau_C^{(1)}=\tau_C^{(2)}=\tau_C$~\cite{PAN12}. First, consider the case that the feedback delays are also equal, $\tau_K^{(1)}=\tau_K^{(2)}=\tau_K$. Then, if $2\tau_C$ and $\tau_K$ are approximately commensurate, \emph{i.e.},
\begin{equation}\label{eq:spiking_condition}
2\tau_CN_C\approx \tau_KN_K\text{ with integers }N_{C,K}\in\mathbb{N}, 
\end{equation}
the dynamics exhibits self-sustained coherent pulse trains with an inter-spike-interval (period) of
\begin{equation}\label{eq:period}
T=\tau_K/N_C=2\tau_C/N_K,
\end{equation}
and a relative phase between the two nodes of $0$ or $\pi$, depending on $N_K$. More specifically, the oscillations are in-phase (anti-phase), if $N_K$ is even (odd).
Coherent spiking is expected even if the coupling delays differ from Eq.~\eqref{eq:spiking_condition} by an amount on the order of the pulse width of the excitable system~\cite{PAN12}.

To test these results experimentally, we realize this coupling scheme with a setup shown in Fig.~\ref{fig:2node_exp}b.
We use $n_{\tau,C}$ and $n_{\tau,K}$ pairs of inverter gates to create the delay lines for coupling and feedback satisfying $\tau_C^{(1)}\approx\tau_C^{(2)}$ and $\tau_K^{(1)}\approx\tau_K^{(2)}$, respectively; thus, the same number of inverter gates are employed in both cases. However, these delays are not exactly equal, because of heterogeneity in the logic gates' propagation delays.

Furthermore, we use two- and three-input logic gates to combine the two delay lines and connections for external stimuli at the input of nodes. For that purpose, we use OR gates, so that any pulse at the input of this logic gate will be passed to the excitable node. However, for larger networks, an $N$-input logic gate that combines $N$ inputs to an excitable node can be defined as desired using a $2^N$-entry look-up table. This so-called synapse (when the excitable node is considered the soma of a silicon neuron) \cite{IND11} allows for implementing inhibitory and excitatory connections and also to vary the coupling strength. The coupling strength is understood as the number of high inputs required for the ``synapse'' to pass on a pulse to the ``soma'' (excitable node).

This setup with delay lines and synapses as described above shows coherent spiking in Fig.~\ref{fig:2node_exp}c,d, when perturbed with a single pulse out of the quiescent state.
The numeric values for the delays satisfy $\tau_C\approx\tau_K\approx22\un{ns}$ ($N_C=1$, $N_K=2$) and $\tau_K\approx2\tau_C\approx44\un{ns}$ ($N_C=1$, $N_K=1$) for Fig.~\ref{fig:2node_exp}c and d, respectively. With these two numerical values, we expect from Eq.~\eqref{eq:spiking_condition} oscillations with period $T$ of $22\un{ns}$ and $44\un{ns}$, respectively. 
This behavior is found approximately in the experiment, where in-phase and anti-phase oscillations are seen with periods of $T=(23.0\pm0.2)\un{ns}$ and $T=(44.8\pm0.2)\un{ns}$, respectively.
For both sets of parameters, we observe small mismatch ($<5\%$) between experiment and theory, likely due to the large processing time $h$.

\section{Model}
In this section, we derive a Boolean map to describe the excitable node theoretically. In contrast to the experimental implementation, this model allows only for Boolean states, \emph{i.e.}, $V\in\left\{V_\mathrm{high},V_\mathrm{low}\right\}$, the low and high voltage of logic gates. 

We model the three components of the setup (Fig.~\ref{fig:setup_artificial_neuron}c), namely the AND gate and the two PGs, separately.
First, we describe the AND gate with output signal $V_\mathrm{AND}^{(j)}(t)$, where the superscript $j$ denotes the nodes in the network. It is modeled by the map
\begin{equation}\label{eq:model}
 V_\mathrm{AND}^{(j)}(t+\Delta) = V_\mathrm{in}^{(j)}(t) \:\wedge \neg V_\mathrm{ref}^{(j)}(t),
\end{equation}
where $\wedge$ and $\neg$ denote the Boolean AND and NOT operations, respectively, and $\Delta$ is the time step of the map. The NOT operation accounts for an inverted input to the AND gate, as shown in the setup.
Second, the PGs denoted by $n_\mathrm{pulse}$ and $n_\mathrm{ref}$ in the setup are modeled by taking the flip-flop and the delay line combined with the XOR gate into account. 
The flip-flop creates events after a positive edge (PE) in $V_\mathrm{AND}^{(j)}$.
The delay line, together with the XOR gate, results in output pulses of the two PGs, \emph{i.e.}, a high voltage for the time intervals $\left[s,s+T_\mathrm{pulse}\right]$ and $\left[s,s+T_\mathrm{ref}\right]$, respectively, after a PE in $V_\mathrm{AND}^{(j)}$ at time $s$ (denoted in the following as $V_\mathrm{AND}^{(j)}(s)=\mathrm{PE}$). Both, the flip-flop and the delay line combined with the XOR gate, are mathematically expressed as
\begin{equation}\label{eq:pulse}
 V_\mathrm{out/ref}^{(j)}(t)=
  \begin{cases} 
   V_\mathrm{high} & \mbox{if } \exists {s\in\left(t-T_\mathrm{pulse/ref},t\right]}:\\    & V_\mathrm{AND}^{(j)}(s)=\text{PE} \\
   V_\mathrm{low} & \mbox{otherwise,} 
  \end{cases}
\end{equation}
for the two PGs denoted in the setup as $n_\mathrm{pulse}$ and $n_\mathrm{ref}$, respectively.
This description does not account for the processing time $h$ of the flip-flop.

\begin{figure}[tb]
\centering
\includegraphics[width=\linewidth]{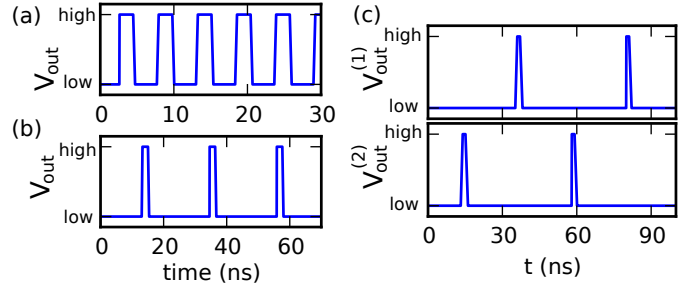}
\caption{(Color online) Simulation of Boolean map for 
$T_\mathrm{pulse}=2.1\un{ns}$, $T_\mathrm{ref}=5.3\un{ns}$.
(a) Output of one node with constant stimulus, corresponding to Fig.~\ref{fig:1node_exp}b. (b) One node with delayed feedback, corresponding to Fig.~\ref{fig:1node_feedback}c with $\tau=21.3\un{ns}$. (c) Two delay-coupled nodes with delayed feedback, corresponding to Fig.~\ref{fig:2node_exp}d with delays $\tau_C=22\un{ns}$, $\tau_K=44\un{ns}$.}
\label{fig:set_model}
\end{figure}

We model the three experiments in this paper with a time step $\Delta=0.01\un{ns}$. For the first experiment of one excitable node with constant input, we set $V_\mathrm{in}=V_\mathrm{high}$ in the model, which results in dynamics shown in Fig.~\ref{fig:set_model}a. The second experiment with delayed feedback of a single node is modeled with $V_\mathrm{in}(t)=V_\mathrm{out}(t-\tau)$ (Fig.~\ref{fig:set_model}b). Due to the delayed feedback, our theoretical description is a Boolean delay equation \cite{GHI85}, which requires an initial history function for initialization. Here, we initialize $V_\mathrm{in}(t)$ with a pulse on the interval $\left[-\tau,0\right]$.
Finally, the third experiment is modeled in Fig.~\ref{fig:set_model}c with 
$V_\mathrm{in}^{(1)}(t)=(V_\mathrm{out}^{(1)}(t-\tau_\mathrm{K}) \vee V_\mathrm{out}^{(2)}(t-\tau_\mathrm{C}))$ and
$V_\mathrm{in}^{(2)}(t)=(V_\mathrm{out}^{(2)}(t-\tau_\mathrm{K}) \vee V_\mathrm{out}^{(1)}(t-\tau_\mathrm{C}))$, where $\vee$ indicates the OR operation.
As above, we initialize the system with a pulse input to one excitable node.

The dynamics generated by the map is similar to the experiment in the overall picture, but in detail the waveforms differ, as the experiment shows imperfections, such as amplitude and timing noise and low-pass filtering effects. To capture these effects, we could use a model based on a set of delay differential equations with stochastic driving terms to describe all logic gates and propagation times in our setup, similar to Refs.~\cite{MES97,GLA98,EDW00}.

\section{Discussion}
Our excitable systems, designed in a bottom-up approach with autonomous Boolean circuits, display dynamics that are not as rich as for system designed in a top-down approach. For example, silicon neurons that are based on analog electronic components are known to show dynamics almost identical to biological neurons \cite{IND11}.
However, our approach allows for large networks, as it relies solely on logic gates. On the single FPGA used in this letter, there are enough logic gates to implement more than $10{,}000$ interconnected excitable systems; but constraints on the wiring placement on the chip may decrease this number depending on the topology.

Furthermore, the logic gates and, hence, also our excitable system operate at time-scales on the order of nanoseconds, which is six to nine orders of magnitude faster than common silicon neurons that operate on a time scale of seconds, and a thousand times faster than the fastest so-called accelerated-time silicon neurons \cite{IND11}.

The fast time scales require us to implement links between excitable systems physically, thereby integrating both the network nodes and their connection on a single off-the-shelf reconfigurable electronic chip. This is another advantage over the VLSI approach (see Ref.~\cite{ART06}) where two chips are needed: a custom analog and a reconfigurable digital chip to implement the silicon neurons on the first and the wiring, the topology, on the latter. For the communication between both chips, an address-event representation is utilized, which leads to discretization errors in the coupling.
By including the entire setup on an FPGA and not using custom chips, our network realization becomes far less expensive than common approaches ($\sim \$300$) and excludes their discretization errors.

Artificial neural networks can be used for bio-inspired data-processing and machine learning. For this so-called reservoir computing, our network's simple spiking dynamics is suitable \cite{sch08k}.

\section{Conclusion}
We have proposed and built excitable nodes that are controllable in pulse width and refractory period. 
A single excitable node responds to a one-time stimulus with a single pulse and, when connected to a time-delayed network, our excitable nodes show self-sustained oscillations with phases and periods that are controlled by the coupling delays.
This dynamics can be reproduced by a simple Boolean model and agrees quantitatively with theoretical studies based on Fitz\-Hugh-Nagumo models, indicating that our setup is suitable for fundamental studies on excitable networks. 

As our setup relies exclusively on asynchronous logic gates and flip-flops, it is fully integrable on an inexpensive off-the-shelf FPGA, allows for thousands of nodes, and operates at speeds a thousand times faster than the fastest silicon neurons. 
Furthermore, it requires only little knowledge about electronic circuits, but mostly knowledge about a so-called hardware description language, which makes our experimental network science approach accessible for biologists and physicists. It will be invaluable as a test-bed for network science and might be applicable to ultra-fast neuro-inspired data processing.

\acknowledgments
D.P.R.,D.R., and D.J.G. gratefully acknowledge the financial support of the U.S. Army Research Office Grants W911NF-11-1-0451 and W911NF-12-1-0099.
E.S. and D.P.R. acknowledge support by the DFG in the framework of SFB~910. We thank Seth Cohen, Kristine Callan, and Thomas Dahms for fruitful discussions.

\end{document}